\documentclass{iau}
\usepackage{graphicx}

\title[NASIM: Advanced LSB reduction for VISTA] 
{The Infrared Glow of Galactic Outskirts: A New Window with NASIM}

\author[Elham Saremi et al.]   
{Elham Saremi$^{1,2,3}$,
          Ignacio Trujillo$^{2,3}$,
          Mohammad Akhlaghi$^4$,
          Zohreh Ghaffari$^{5,6}$,
          Johan H. Knapen$^{2,3}$,
          Manda Banerji$^1$,
          Helmut Dannerbauer$^{2,3}$,
          \and
          Sébastien Comerón$^{3,2}$}

\affiliation{$^1$School of Physics \& Astronomy, University of Southampton, Highfield Campus, Southampton SO17 1BJ, UK \\ email: {\tt e.saremi@soton.ac.uk} \\[\affilskip]
$^2$Instituto de Astrofísica de Canarias, C/ Vía Láctea
              s/n, 38205 La Laguna, Tenerife, Spain \\
$^3$Departamento de Astrofísica, Universidad de La Laguna, 38205 La Laguna, Tenerife, Spain \\
$^4$Centro de Estudios de Física del Cosmos de Aragón (CEFCA), Plaza San Juan 1, 44001 Teruel, Spain \\
$^5$INAF-Osservatorio Astronomico di Trieste, Via G. B. Tiepolo 11, 34143 Trieste, Italy \\
$^6$IFPU, Institute for Fundamental Physics of the Universe, via Beirut 2, 34151 Trieste, Italy}

\pubyear{2025}
\volume{403}  
\pagerange{119--126}
\jname{The Hidden Beauty of the Galactic Outskirts}
\editors{A.C. Editor, B.D. Editor \& C.E. Editor, eds.}
\begin{document}

\maketitle

\begin{abstract}
The outskirts of galaxies retain fossil records of their assembly history and are dominated by old stellar populations shaped by tidal interactions, accretion events, and environmental processes. Observing these low-surface-brightness (LSB) regions in the near-infrared (NIR), particularly in the $K_{\rm s}$ band, is crucial for tracing evolved stars and minimising the effects of dust. Yet, ground-based NIR studies of faint extended structures remain challenging because of the bright and variable sky background and complex detector systematics.

To overcome these limitations, we developed NASIM, the Near-infrared Automated low Surface brightness reduction In Maneage pipeline, a fully automated and reproducible framework designed to recover faint structures in VISTA/VIRCAM data. Using GNU Astronomy Utilities (Gnuastro), NASIM corrects large-scale flat-field patterns and instrumental signatures across the VIRCAM detectors while preserving LSB emission that can be strongly suppressed in conventional NIR reductions.

We apply NASIM to $K_{\rm s}$-band imaging of the Euclid Deep Field South through the KEDFS programme, a key region also targeted by Rubin/LSST and Euclid. The final reduction reaches a surface-brightness limit of $\mu_{K_{\rm s}}\simeq27.7$~mag~arcsec$^{-2}$, measured at 3$\sigma$ over 100~arcsec$^2$, approximately 67 times deeper than 2MASS and 11 times deeper than VHS in surface-brightness sensitivity. Direct comparisons with publicly available VISTA products show that NASIM preserves diffuse emission without substantially compromising compact-source depth. These results open a deep NIR view of faint galaxy outskirts, LSB galaxies, and intracluster light, providing a reproducible ground-based foundation for future studies of the faint Universe.


\end{abstract}

\section{Introduction}

The low-surface-brightness (LSB) Universe remains poorly explored in the near-infrared (NIR) from the ground, mainly because the bright atmospheric background makes the detection of faint extended structures extremely challenging. In the $K_{\rm s}$ band, the sky is much brighter than the optical sky ($\mu_K\simeq14.85$~mag~arcsec$^{-2}$; \cite[Cuby et al. 2000]{Cuby2000}), producing substantial noise that hinders the recovery of diffuse emission such as galaxy outskirts, LSB dwarfs, tidal structures, and intracluster light (ICL). These difficulties are particularly severe for wide-field surveys, where the sky varies both spatially and temporally (\cite[Borlaff et al. 2019]{Borlaff2019}). Despite these challenges, the NIR is essential for galaxy-evolution studies because it traces the old stellar populations that dominate galaxy mass (\cite[Zibetti et al. 2009]{Zibetti2009}, \cite[Conroy 2013]{Conroy2013}). In particular, $K_{\rm s}$-band light is relatively insensitive to dust attenuation and recent star formation, making it a powerful tracer of stellar mass in galaxy outskirts and diffuse systems (e.g., \cite[Kochanek et al. 2001]{Kochanek2001}, and \cite[McGaugh \& Schombert 2014]{McGaugh2014}).

VISTA/VIRCAM has produced a rich legacy of southern-sky NIR imaging, with particular importance in the $K_{\rm s}$ band. The standard processing pipelines, such as CASU (Cambridge Astronomy Survey Unit; \cite[Irwin et al. 2004]{Irwin2004}, \cite[Emerson et al. 2006]{Emerson2006}, and \cite[Cross et al. 2012]{Cross2012}), are highly effective for point-source and compact-galaxy science. However, sky subtraction techniques and source detection algorithms, optimised for compact sources, can inadvertently suppress or remove extended emission (e.g., \cite[Fliri \& Trujillo 2016]{Fliri2016}, and \cite[Rom\'an et al. 2021]{Roman2021}). 

To address this limitation, we developed NASIM, the Near-infrared Automated low Surface brightness reduction In Maneage, a dedicated and reproducible reduction pipeline for VISTA/VIRCAM data (\cite[Saremi et al. 2025]{Saremi2025}). NASIM is designed to remove large-scale instrumental signatures and flat-field patterns associated with the complex VIRCAM detector layout while preserving faint extended emission.

\section{NASIM}

NASIM is an LSB-oriented reduction pipeline developed for VISTA/VIRCAM data, where the central challenge is to remove the dominant NIR background and detector-level systematics without suppressing real diffuse emission. The pipeline is fully reproducible through Maneage (Managing data lineage; \cite[Akhlaghi et al. 2021]{Akhlaghi2021}). In this way, all stages of the reduction, from input data to final stacks, are controlled through a transparent and repeatable workflow. NASIM also uses GNU Astronomy Utilities (Gnuastro; \cite[Akhlaghi \& Ichikawa 2015]{AkhlaghiIchikawa2015}, \cite[Akhlaghi 2019]{Akhlaghi2019}) for the main image-processing steps.

The main NASIM reduction workflow consists of data preparation, dark-signal removal, sky flat-fielding, correction of detector-level noise artefacts, astrometric calibration, sky subtraction, image weighting, photometric calibration, resampling, and final stacking. The technical details of these steps are presented in \cite[Saremi et al. (2025)]{Saremi2025}. Here, we focus on three aspects of the workflow that are most directly connected to the recovery of LSB emission. \\

{\underline{\it Adaptive flat-fielding}}. We use a novel flat-fielding strategy in NASIM, referred to as the adaptive flat, in which calibration frames are built directly from the science exposures. The input images used for this process are selected separately for each exposure, allowing the pipeline to construct a unique master flat for every image. This procedure is implemented with Gnuastro tools and is designed to remove large-scale instrumental patterns and correct the complex detector-level structures across the VIRCAM array.

{\underline{\it Conservative sky subtraction}}. We apply a conservative sky-subtraction strategy that avoids overfitting large-scale background variations and helps preserve genuine diffuse emission. We combine this with optimised object masking using \textsc{NoiseChisel} (\cite[Akhlaghi \& Ichikawa 2015]{AkhlaghiIchikawa2015}, \cite[Akhlaghi 2019]{Akhlaghi2019}) and careful sky modelling on appropriate spatial scales, reducing biases in the reconstruction of faint extended structures.

{\underline{\it Robust stacking}}. We perform robust outlier rejection during the final coaddition and apply a weighted stacking scheme that accounts for image quality and background noise. This strategy suppresses residual artefacts, detector defects, and transient features, while preserving extended LSB signal across multiple exposures.

\section{Data}

The primary dataset used here is the $K_{\rm s}$-band imaging of the Euclid Deep Field South obtained with VISTA/VIRCAM through the KEDFS programme (ID: 106.21LU.001; PI: Nonino, M.). KEDFS covers approximately 20 deg$^2$ around R.A.\,(J2000) = 04$^{\mathrm{h}}$\,04$^{\mathrm{m}}$\,57$^{\mathrm{s}}$.84 and Dec\,(J2000) = $-48^\circ$\,25$'$\,22$''$.8, close to the South Ecliptic Pole. The field is particularly valuable because it combines deep $K_{\rm s}$-band VISTA imaging with multiwavelength coverage, including Euclid VIS, $Y$, $J$, and $H$ observations (\cite[Euclid Collaboration et al. 2023]{Euclid23}), mid-infrared data from the Spitzer Space Telescope (\cite[Euclid Collaboration et al. 2022]{Euclid22}), and forthcoming optical coverage from Rubin/LSST (\cite[Gris et al. 2024]{Gris24}). Its location also lies within the continuous viewing zones of both Roman and JWST, making KEDFS a valuable legacy field for studies of galaxy evolution and faint diffuse structures. The observations consist of 15 VIRCAM tiles, with a total integration time of 207~hours collected between December 2020 and November 2022. Further details of the survey, reduction, and released products are given in \cite[Saremi et al. (2025)]{Saremi2025}.

For comparison with a standard VISTA reduction, we also use data from the VISTA Deep Extragalactic Observations (VIDEO) survey (\cite[Jarvis et al. 2013]{Jarvis2013}). VIDEO provides deep multi-band VISTA imaging over extragalactic fields, including XMM-LSS. In this work, the VIDEO data are used to compare the publicly available VIDEO reduction with the same field processed by NASIM, providing a direct test of how different reduction strategies affect the recovery of extended LSB emission.

\section{Revealing the faint NIR Universe with NASIM} 

We demonstrate the impact of NASIM through two complementary applications. First, we compare NASIM reductions with publicly available VIDEO products in the XMM-LSS field, using nearby galaxies to test how the reduction strategy affects the recovery of extended emission, the apparent morphology, and the measured $K_{\rm s}$-band surface-brightness profiles. We then turn to the KEDFS field, where the same LSB-optimised approach reaches the depth and background stability required to study diffuse stellar structures in the Euclid Deep Field South, from LSB galaxies to ICL.

\subsection{An LSB view beyond conventional reductions}

To illustrate the practical advantages of this LSB-optimised reduction scheme, we compare NASIM with the publicly available reduced products from the VIDEO survey. The VIDEO reduction is primarily optimised for compact-source detection and catalogue production: the sky background is estimated from several unaligned, jittered pawprints, with the target frame excluded to reduce self-subtraction, and masks are applied mainly to the brightest sources (\cite[Jarvis et al. 2013]{Jarvis2013}). This strategy provides deep and valuable survey products for faint compact objects, but diffuse extended components can still be affected by systematic over-subtraction in the LSB regime.

\begin{figure}[t]
\centering
\includegraphics[width=\textwidth]{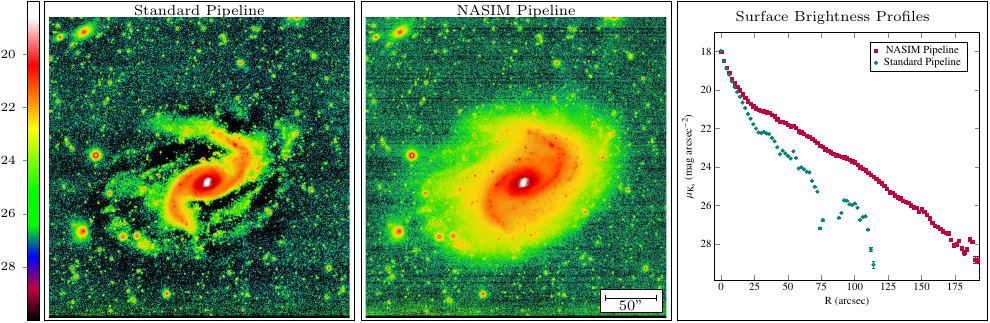}
\caption{LSB view of NGC~895 in the publicly available VIDEO reduction and in the NASIM reduction. The left and middle panels show the two reductions of the same VISTA/VIRCAM data, respectively; the colour bar gives surface brightness in mag~arcsec$^{-2}$. NASIM preserves a more extended and coherent LSB structure around the galaxy. The right panel shows the corresponding $K_{\rm s}$-band surface-brightness profiles, demonstrating that the NASIM reduction traces the galaxy light to larger radii and fainter levels.}
\label{fig:ngc895}
\end{figure}

We select the VIDEO XMM-LSS field for this comparison. Figure~\ref{fig:ngc895} shows a side-by-side comparison between the publicly available VIDEO image and the NASIM-processed image for the nearby galaxy NGC~895. The NASIM reduction reveals a more coherent and extended LSB component around the galaxy. This morphological difference is confirmed quantitatively by the radial $K_{\rm s}$-band surface-brightness profiles shown in the right panel of the same figure. The NASIM profile extends smoothly to larger radii and fainter surface-brightness levels, indicating a more complete recovery of the galaxy outskirts.

\begin{figure}[b]
\centering
\includegraphics[width=\textwidth]{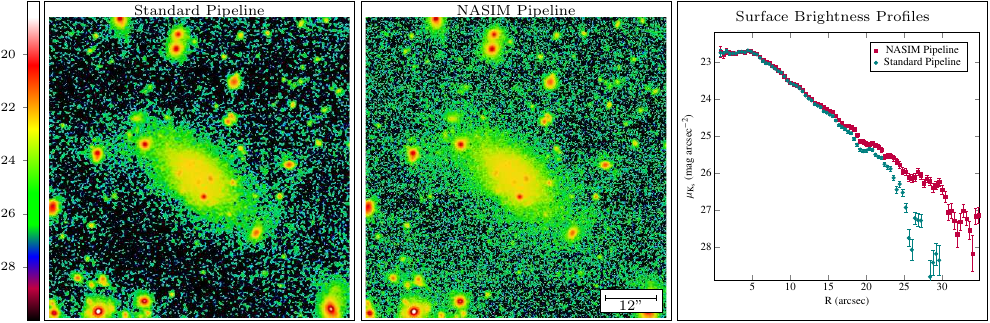}
\caption{Recovery of LSB emission in the nearby dwarf galaxy LEDA~3098171. The left and middle panels show the publicly available VIDEO reduction and the NASIM reduction of the same VISTA/VIRCAM data, respectively; the colour bar gives surface brightness in mag~arcsec$^{-2}$. While the central body is detected in both images, the faint outer envelope is more clearly preserved by NASIM. The right panel shows the corresponding $K_{\rm s}$-band surface-brightness profiles, demonstrating that the diffuse component remains measurable to fainter levels in the NASIM reduction.}
\label{fig:dwarf}
\end{figure}

A second example is shown in Figure~\ref{fig:dwarf} for the nearby dwarf galaxy LEDA~3098171. This object provides an important test because its LSB component is smaller and fainter than that of NGC~895. Both reductions detect the central body, but the NASIM-processed data preserve a faint outer envelope that is much less apparent in the public VIDEO product. Recovering this extended component is important for measuring the total light of dwarf galaxies and therefore for deriving more reliable sizes, luminosities, and stellar masses.

The conservative sky-subtraction strategy used in NASIM is designed to preserve LSB emission without substantially compromising the depth of the final image. In 2~arcsec diameter apertures, the 5$\sigma$ limiting magnitude is 23.67~mag for the NASIM reduction and 23.72~mag for the VIDEO reduction. This small difference shows that NASIM maintains competitive sensitivity to faint compact sources, while also enabling the robust recovery of diffuse LSB structures.

\subsection{LSB science cases in KEDFS}

The KEDFS survey provides a new, deep $K_{\rm s}$-band view of the Euclid Deep Field South. In this section, we focus on the LSB science enabled by applying NASIM to this field. The final stacks reach a surface-brightness limit of $\mu_{K_{\rm s}}\simeq27.7$~mag~arcsec$^{-2}$, measured at 3$\sigma$ over 100~arcsec$^2$, corresponding to a sensitivity approximately 67 times deeper than 2MASS and 11 times deeper than VHS (\cite[Saremi et al. 2025]{Saremi2025}). This depth places KEDFS in a NIR regime that has previously been difficult to access over a wide field, enabling the study of diffuse stellar structures directly in the $K_{\rm s}$ band. The combination of KEDFS depth and NASIM reduction makes this dataset particularly well suited to studying low-contrast stellar structures, from galaxy outskirts and diffuse dwarf systems to the intracluster component in galaxy clusters.

\begin{figure}
\centering
\begin{minipage}{0.49\textwidth}
    \centering
    \includegraphics[width=\textwidth]{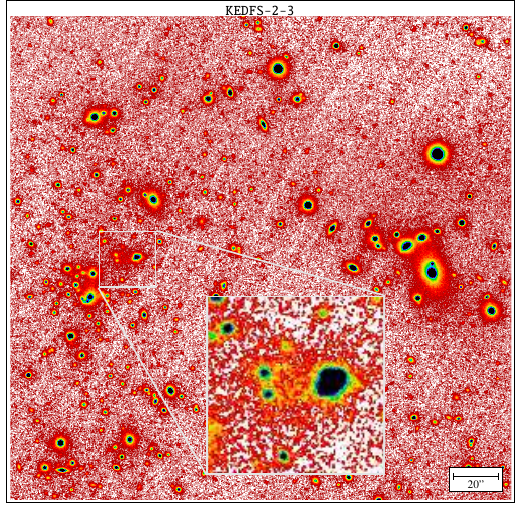}
\end{minipage}
\hfill
\begin{minipage}{0.49\textwidth}
    \centering
    \includegraphics[width=\textwidth]{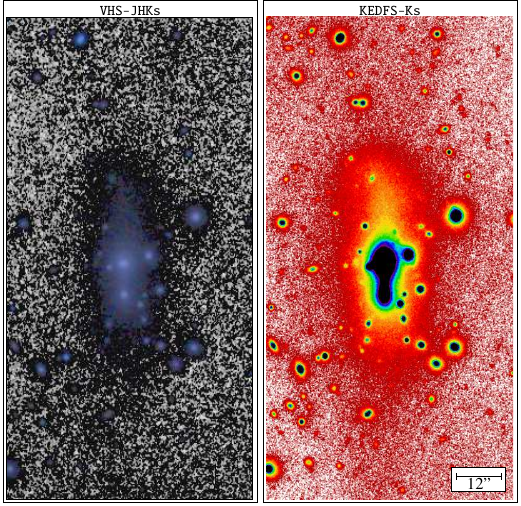}
\end{minipage}
\caption{Examples of LSB structures recovered from the KEDFS $K_{\rm s}$-band data. Left: UDG candidate SMDG~J0406539--480442, with a zoom-in panel highlighting its diffuse stellar body. Right: comparison between a VHS $JHK_{\rm s}$ colour image and the deeper NASIM $K_{\rm s}$ reduction of the galaxy cluster ACT-CL~J0414.2--4612. The deeper NASIM image reveals extended ICL around the brightest cluster galaxy that is much less apparent in the shallower VHS data.}
\label{fig:kedfs_examples}
\end{figure}

LSB galaxies provide a natural science case for the NASIM-processed KEDFS data. Their diffuse stellar emission is intrinsically difficult to detect and characterise, because the low-contrast signal can be close to the residual background level. Within this broad population, ultra-diffuse galaxies (UDGs) occupy an especially important regime: with central surface brightnesses of $\mu_g \sim 24$--$26$~mag~arcsec$^{-2}$ and effective radii of $r_{\rm e} \sim 1.5$--$4.5$~kpc, they are often interpreted as promising systems for probing low-density, dark-matter-dominated galaxies (e.g., \cite[van Dokkum et al. 2015]{vanDokkum15}; \cite[Zaritsky et al. 2023]{Zaritsky23}). Although most UDG discoveries have relied on deep optical imaging, $K_{\rm s}$-band data offer a more direct tracer of the old stellar populations that dominate their stellar mass. The left panel of Figure~\ref{fig:kedfs_examples} shows the UDG candidate SMDG~J0406539--480442 in the KEDFS field (\cite[Zaritsky et al. 2019]{Zaritsky19}), illustrating that NASIM can recover such diffuse systems directly in the NIR.

As another key component of the LSB Universe, the ICL provides a direct tracer of the assembly history of galaxy clusters, since it is produced by stars stripped from galaxies through tidal interactions, mergers, and other dynamical processes within the cluster environment (\cite[Montes \& Trujillo 2014]{Montes2014}). The right panel of Figure~\ref{fig:kedfs_examples} compares the shallower VHS $JHK_{\rm s}$ colour image with the deeper NASIM $K_{\rm s}$ reduction of the cluster ACT-CL~J0414.2--4612. The colour composite was produced using the \textsc{astscript-color-faint-gray} script from Gnuastro, which applies a non-linear transformation to compress the dynamic range of bright sources and enhance faint diffuse structures (\cite[Infante-Sainz \& Akhlaghi 2024]{InfanteSainz2024}). In the NASIM reduction, the ICL is revealed as a diffuse, spatially coherent component surrounding the brightest cluster galaxy, while it remains much less apparent in the VHS composite. This comparison highlights the importance of both depth and LSB-optimised reduction for NIR studies of ICL. \\

Looking forward, NASIM provides a foundation for future LSB studies in the NIR, especially in synergy with major space-based surveys such as Euclid, Roman, and ARRAKIHS, for which deep ground-based imaging will remain an essential complement. More broadly, this work highlights the importance of transparent and fully reproducible reduction workflows for obtaining reliable measurements of faint diffuse emission in the era of large astronomical surveys.

\section*{Acknowledgements}
{\footnotesize
We acknowledge grants PID2022-136505NB-I00, PID2022-140869NB-I00, PID2022-143243NB-I00, PID2023-149139NB-I00, PID2021-124918NA-C43, and PID2024-162229NB-I00, funded by MCIN/AEI/10.13039/501100011033 and EU, ERDF.}

\end{document}